\documentclass{vietnam}

\usepackage{subfigure} 

\def\mco{\multicolumn}

\def\be{\begin{equation}}
\def\ee{\end{equation}}

\newcommand{\barray}{\begin{eqnarray}}
\newcommand{\earray}{\end{eqnarray}}
\newcommand{\nn}{\nonumber \\}

\newcommand{\beq}{\begin{equation}}
\newcommand{\eeq}{\end{equation}}
\newcommand{\ba}{\begin{array}}
\newcommand{\ea}{\end{array}}
\newcommand{\bea}{\begin{eqnarray}}
\newcommand{\eea}{\end{eqnarray} }
\newcommand{\bal}{\begin{align}}
\newcommand{\eal}{\end{align}}
\newcommand{\bi}{\begin{itemize}}
\newcommand{\ei}{\end{itemize}}
\newcommand{\ben}{\begin{enumerate}}
\newcommand{\een}{\end{enumerate}}
\newcommand{\bc}{\begin{center}}
\newcommand{\ec}{\end{center}}
\newcommand{\bt}{\begin{table}}
\newcommand{\et}{\end{table}}
\newcommand{\btb}{\begin{tabular}}
\newcommand{\etb}{\end{tabular}}

\def\cl{{\mathcal L}}

\def\pa{\partial}

\def\hc{{\rm h.c.}}

\newcommand{\Photo}{\includegraphics[height=35mm]{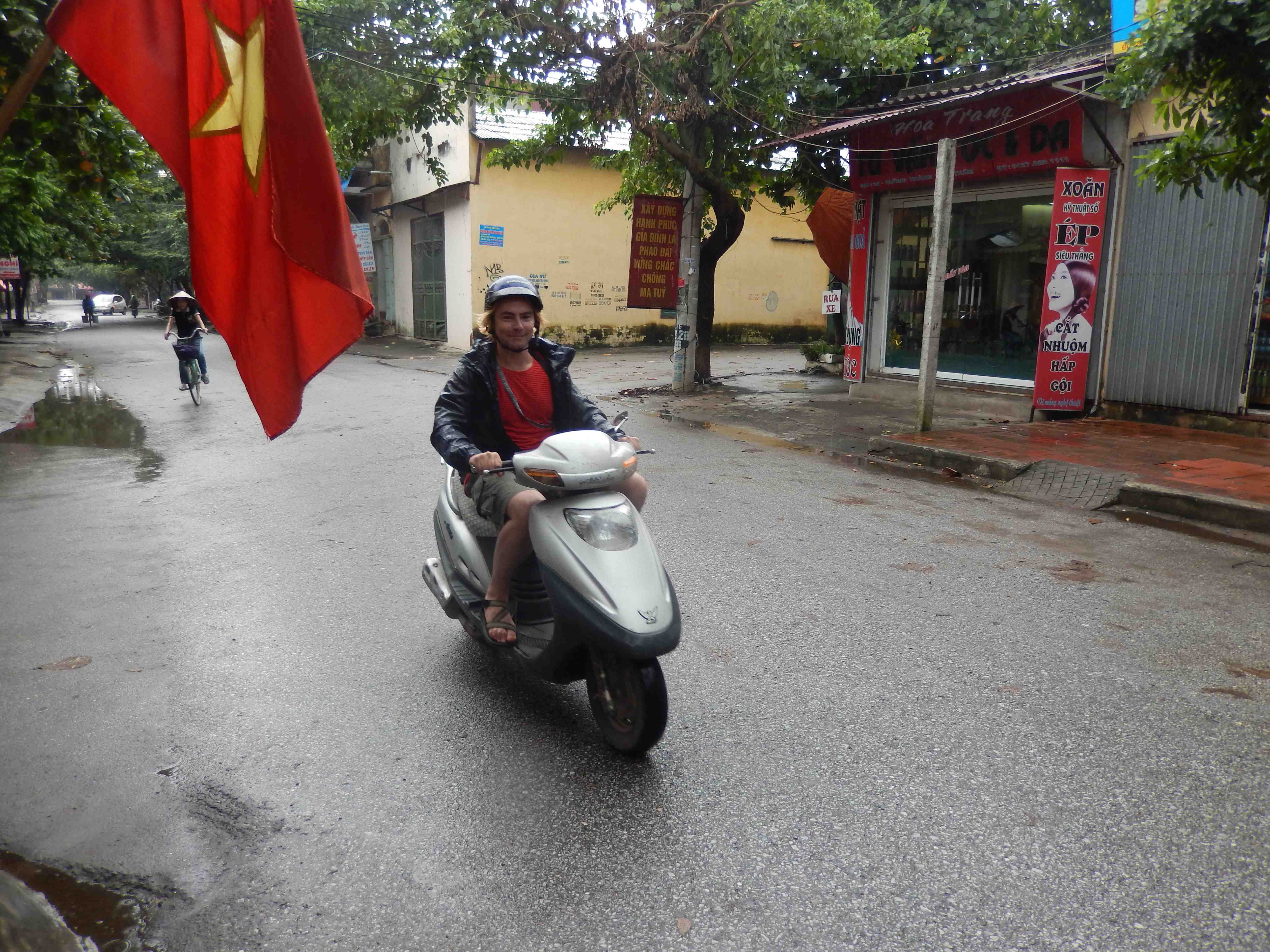}}

\begin{document}
\vspace*{4cm}
\title{Constraints on Higgs Couplings and Physics Beyond the Standard Model}

\author{Herm\`es B\'elusca-Ma\"ito, Adam Falkowski}

\address{Laboratoire de Physique Th\'eorique, CNRS -- UMR 8627, \\ 
B\^at. 210 Universit\'e de Paris-Sud 11 F-91405 Orsay Cedex France \\ 
. }

\vspace{0.5cm}

\maketitle
\abstracts{
We summarize current constraints on the couplings of the Higgs boson in the framework of an effective theory beyond the Standard Model.  
{\it Article prepared for the proceedings of the 9th Rencontres du Vietnam ``Windows on the Universe" Quy Nhon Vietnam, August 11-17 2013.}
}

\section{Introduction}

The discovery of the Higgs boson at the LHC~\cite{Aad:2012tfa} was a spectacular confirmation of the central prediction of the Standard Model (SM).
Nevertheless, it is possible that precision studies of the Higgs boson 
will reveal new physics beyond the SM. 
The {\em effective theory} framework offers a model-independent approach to address this issue. 
The underlying assumption is that there are no new particles (beyond those of the SM) with masses near the weak scale.
In this talk we present up-to-date constraints on the parameters of the leading effective theory operators that govern the Higgs couplings to matter.

\section{Effective Lagrangian for Higgs interactions}

We assume that the Higgs boson $h$ is a part of the Higgs field $H$ that transforms as the $({\bf 1}, {\bf 2})_{1/2}$ representation under the SM $SU(3)_C \times SU(2)_L \times U(1)_Y$ gauge group and obtains an expectation value $v$. 
Then one can organize the effective Lagrangian as an expansion 
$\cl_{\rm eff} = \cl_{\rm SM} + \cl_{D=5} + \cl_{D=6} + \dots$~, where each term consists of gauge invariant local operators of canonical dimension $D$ constructed out of the SM fields. 
The leading term is the SM Lagrangian which contains operators up to dimension 4.
The only operators at dimension 5 are of the form $(LH)^2$; they give masses to neutrinos but have no observable impact on Higgs phenomenology.
At dimension 6, the minimal non-redundant set of operators was given by Grzadkowski {\it et al.}~\cite{Grzadkowski:2010es}; we use the equivalent basis written down in Contino {\it et al.}~\cite{Contino:2013kra}. We assume $\cl_{D=6}$ contains no new sources of flavor, CP and baryon number violation.
Then the dimension 6 operators lead to the following couplings of a single Higgs boson to pairs of SM fields:
\bea
\label{eq:leff}
\cl_{h} &=& {h \over v} \left (
2 c_{V} m_W^2 W_\mu^+ W_\mu^ - + c_{V,Z} m_Z^2 Z_\mu Z_\mu 
 - c_u \sum_{f = u,c,t} m_f \bar f f - c_d \sum_{f = d,s,b} m_f \bar f f - c_l \sum_{f = e,\mu,\tau} m_f \bar f f 
\right . \nn && \left .
+ {1 \over 4} c_{gg} G_{\mu\nu}^a G_{\mu\nu}^a 
- {1 \over 2} c_{WW} W_{\mu\nu}^+ W_{\mu\nu}^-
- {1 \over 4} c_{\gamma \gamma} \gamma_{\mu\nu} \gamma_{\mu\nu}
- {1 \over 4} c_{ZZ} Z_{\mu \nu} Z_{\mu \nu}
- {1 \over 2} c_{Z \gamma} \gamma_{\mu\nu} Z_{\mu\nu}
\right . \nn && \left . 
+ \kappa_{Z \gamma} \pa_\nu \gamma_{\mu \nu} Z_\mu 
+ \kappa_{Z Z} \pa_\nu Z_{\mu \nu} Z_\mu 
+ ( \kappa_{WW} \pa_\nu W_{\mu \nu}^+ W_\mu^- + \hc )
\vphantom{\sum_{f}}\right ),
\eea
$\cl_{D=6}$ contains also the so-called vertex and dipole operators that modify Higgs couplings to 3 or more SM fields, but we ignore them here.
The 13 real couplings in Eq.~(\ref{eq:leff}) map to 11 operators in the dimension 6 Lagrangian (one constraint on $c_{ii}$ and one on $\kappa_{ii}$ follow from an accidental custodial symmetry in $\cl_{D=6}$~\cite{Contino:2013kra}).
Given the current precision of experiment and theoretical predictions, the effective operators of dimension greater than 6 are not relevant.

It would be desirable to obtain constraints on all these coefficients using the Higgs data.
However, the data publicly available so far leave important degeneracies, in particular they have a very limiting power of discriminating between different tensor structures of the Higgs coupling to vector bosons.
Therefore we make further assumptions demanding that the combinations of couplings leading to power-divergent corrections to electroweak precision observables vanish.
This leads to the constraints~\cite{Falkowski:2013dza}
\beq
\label{eq:custodial}
c_{V,Z} = c_V,
\quad
c_{WW} = c_{\gamma\gamma} + {g_L \over g_Y} c_{Z\gamma},
\quad 
c_{ZZ} = c_{\gamma\gamma} + {g_L^2 - g_Y^2 \over g_L g_Y} c_{Z\gamma},
\quad
\kappa_{Z\gamma} = \kappa_{WW} = \kappa_{ZZ} = 0, 
\eeq
where $g_L$ and $g_Y$ are the gauge couplings of $SU(2)_L \times U(1)_Y$. 
Only 2 combinations of these constraints follow automatically from $\cl_{D=6}$. 
The remaining ones have to be imposed by hand and represent an assumption about the underlying UV theory.
In the following we work with the effective Lagrangian of Eq.~(\ref{eq:leff}) subject to the constraints of Eq.~(\ref{eq:custodial}).
The Higgs couplings to matter thus depend on 7 free parameters:
\beq
\label{eq:parameters}
c_V, \quad c_u, \quad c_d, \quad c_l, \quad c_{gg}, \quad c_{\gamma \gamma}, \quad c_{Z\gamma}. 
\eeq
The SM Higgs is the limiting case where $c_V = c_{f=u,d,l} =1$ and $c_{gg}=c_{\gamma \gamma} = c_{Z \gamma } = 0$.
Moving away from the SM point, one effect is that the partial decay widths of the Higgs boson are modified. 
For $m_h=125.6$~GeV, the decay widths relative to the SM ones can be expressed in terms of the parameters in Eq.~\ref{eq:parameters} as 
\beq
{\Gamma_{cc} \over \Gamma_{cc}^{\rm SM}} \simeq |c_u|^2, \qquad 
{\Gamma_{bb} \over \Gamma_{bb}^{\rm SM}} \simeq |c_d|^2, 
\qquad 
{\Gamma_{\tau \tau} \over \Gamma_{\tau \tau}^{\rm SM}} \simeq |c_l|^2, 
\eeq
\beq
{\Gamma_{ZZ^* \to 4l} \over \Gamma_{ZZ^* \to 4l}^{\rm SM}} \simeq 
c_V^2 + 0.36\,c_V c_{Z\gamma} + 0.26\,c_Vc_{\gamma\gamma}, 
\quad 
{\Gamma_{WW^* \to 2l2\nu} \over \Gamma_{WW^*\to 2l2\nu}^{\rm SM}} \simeq 
c_V^2 + 0.73\,c_V c_{Z\gamma} + 0.38\,c_V c_{\gamma\gamma} ,
\eeq 
\bea
{ \Gamma_{gg} \over \Gamma_{g g}^{\rm SM}} &\simeq & {|\hat c_{gg}|^2 \over |\hat c_{gg,\rm SM}|^2},
\quad \hat c_{gg} = c_{gg} + 10^{-2} \left [1.28\,c_u- (0.07 - 0.1\,i)\,c_d \right ], \quad |\hat c_{gg,\rm SM}| \simeq 0.012, 
\nn 
{\Gamma_{\gamma \gamma} \over \Gamma_{\gamma \gamma}^{\rm SM}} &\simeq& {|\hat c_{\gamma \gamma}|^2\over |\hat c_{\gamma \gamma,\rm SM}|^2}, 
\quad \hat c_{\gamma \gamma} = c_{\gamma \gamma} + 10^{-2} \left (0.97\,c_V - 0.21\,c_u \right ), \quad |\hat c_{\gamma \gamma,\rm SM}| \simeq 0.0076,
\nn 
{\Gamma_{Z \gamma} \over \Gamma_{Z \gamma}^{\rm SM}} &\simeq& {|\hat c_{Z \gamma}|^2\over |\hat c_{Z \gamma,\rm SM}|^2}, 
\quad \hat c_{Z \gamma} = c_{Z \gamma} + 10^{-2} \left (1.49\,c_V - 0.09\,c_u \right ), \quad |\hat c_{Z \gamma,\rm SM}| \simeq 0.014. 
\eea
The relative branching fraction is given by
${{\rm Br}(h \to XX) \over {\rm Br}(h \to XX)_{\rm SM}} = {\Gamma_{XX} \over \Gamma_{XX, \rm SM}} {\Gamma_{\rm tot, SM} \over \Gamma_{\rm tot}} $, where $\Gamma_{\rm tot}$ is the sum of all partial decay widths. 
Furthermore, the Higgs production cross-section via the gluon fusion $g g \to h $ (ggH), top pair associated production $g g \to h t \bar t$ (ttH),
vector boson fusion $q q \to h qq$ (VBF), and vector boson associated production $q \bar q \to h W/Z$ (VH) are modified as
\beq
{\sigma_{ggH} \over \sigma_{ggH}^{\rm SM}} \simeq {|\hat c_{gg}|^2 \over |\hat c_{gg,\rm SM}|^2 }, 
\quad
{\sigma_{ttH} \over \sigma_{ttH}^{\rm SM}} \simeq c_u^2, 
\quad 
{\sigma_{VBF} \over \sigma_{VBF}^{\rm SM}} \simeq c_V^2 + 0.6\,c_V c_{Z \gamma} + 0.3\,c_V c_{\gamma \gamma}, 
\eeq
\beq
\label{eq:vhrates}
{\sigma_{W H} \over \sigma_{WH}^{\rm SM}} \simeq c_V^2 - 7.0\,c_V c_{Z \gamma} - 3.6\,c_V c_{\gamma \gamma}, 
\quad 
{\sigma_{Z H} \over \sigma_{Z H}^{\rm SM}} \simeq c_V^2 - 5.7\,c_V c_{Z \gamma} - 3.4\, c_V c_{\gamma \gamma}. 
\eeq
The LHC collaborations typically quote the relative rate
$\hat \mu_{XX}^{YH} = {\sigma_{Y H} \over \sigma_{YH}^{\rm SM}}{ {\rm Br}(h \to XX) \over {\rm Br}(h \to XX)_{\rm SM}}$ in a number of channels.
Comparing the measured rates with the SM predictions we can constrain the parameters of the effective Lagrangian.
We will show that the current Higgs and electroweak precision data constrain in a non-trivial way {\em all} the 7 parameters in Eq.~\ref{eq:parameters}.

\section{Experimental data}

The LHC Higgs data we included in our fit are collected in Table~1.
We used 2-dimensional (2D) likelihoods in the $\hat\mu_{ggH+ttH}/\hat\mu_{VBF+VH}$ plane whenever those are provided by experiments.
In these cases the value of $\hat \mu$ is given for illustration only; in the fit we always use the full 2D likelihood function which captures non-trivial correlations between the rates measured for various production modes.
In the channels where only 95\%~CL limits are given we reconstruct $\hat \mu$ assuming the errors are Gaussian.
We also include the combined Tevatron measurements~\cite{Aaltonen:2013kxa}: $\hat{\mu}_{\gamma \gamma}^{\rm incl.}= 6.2^{+3.2}_{-3.2}$, $\hat{\mu}_{WW}^{\rm incl.}= 0.9_{-0.8}^{+0.9}$, 
$\hat{\mu}_{bb}^{VH}= 1.62^{+0.77}_{-0.77}$, $\hat{\mu}_{\tau\tau}^{\rm incl.}= 2.1^{+2.2}_{-2.0}$.
Furthermore, we use electroweak precision measurements from LEP, SLC, and Tevatron collected in Table~1 of Falkowski {\it et al.}~\cite{Falkowski:2013dza}.
To evaluate logarithmically divergent corrections from the Higgs loops to electroweak precision observables we assume $\Lambda = 3$~TeV.

\begin{table}[t]
\caption{The LHC Higgs search results used in our fit.}
\label{tab:exp}
\vspace{0.4cm}
\begin{center}
\renewcommand*{\arraystretch}{1.2} 
\begin{tabular}{|c|l|c|c|}
\hline
\mco{4}{|c|}{\bf ATLAS} \\ \hline 
Production & Decay & $\hat \mu$ & Ref. 
\\ \hline 
2D &
$\gamma \gamma$ & $1.55^{+0.33}_{-0.29}$ & \cite{ATLAS_coup,ATLAS-data-HGaGa}
\\ \cline{2-4} &
$Z Z$ & $1.41^{+0.42}_{-0.33}$ & \cite{ATLAS_coup,ATLAS-data-HZZ}
\\ \cline{2-4} &
$W W$ & $0.98^{+0.33}_{-0.26}$ & \cite{ATLAS_coup,ATLAS-data-HWW}
\\ \cline{2-4} &
$\tau \tau$ & $0.65^{+0.70}_{-0.62}$ & \cite{ATLAS_coup2} 
\\ \hline
VH &
$b b$ & $0.2^{+0.7}_{-0.6}$ & \cite{ATLAS_BB}
\\ \hline
ttH &
$b b$ & $2.69 \pm 5.53$ & \cite{ATLAS_tthBB}
\\ \cline{2-4} &
$\gamma \gamma$ & $-1.39 \pm 3.18$ & \cite{ATLAS_tthGaGa}
\\ \hline 
inclusive &
$Z \gamma$ & $2.96 \pm 6.69$ & \cite{ATLAS_ZGa} 
\\ \cline{2-4} &
$\mu \mu$ & $1.75 \pm 4.26$ & \cite{ATLAS_mumu} 
\\ \hline 
\end{tabular}
\quad
\begin{tabular}{|c|l|c|c|}
\hline
\mco{4}{|c|}{\bf CMS} \\ \hline 
Production & Decay & $\hat \mu$ & Ref. 
\\ \hline 
2D &
$\gamma \gamma$ & $0.77^{+0.29}_{-0.26}$ & \cite{CMS_GaGa}
\\ \cline{2-4} &
$Z Z$ & $0.92^{+0.29}_{-0.24}$ & \cite{CMS_Comb} 
\\ \cline{2-4} &
$W W$ & $0.68^{+0.21}_{-0.19}$ & \cite{CMS_Comb}
\\ \cline{2-4} &
$\tau \tau$ & $1.11^{+0.43}_{-0.41}$ & \cite{CMS_Comb} 
\\ \hline
VH &
$b b$ & $1.00 \pm 0.49$ & \cite{CMS_VHbb}
\\ \hline
VBF &
$b b$ & $0.7 \pm 1.4$ & \cite{CMS_VBFbb}
\\ \hline
ttH &
$b b$ & $1.0^{+1.9}_{-2.0} $ & \cite{CMS_tthComb} 
\\ \cline{2-4} &
$\gamma \gamma$ & $-0.2^{+2.4}_{-1.9}$ & \cite{CMS_tthComb}
\\ \cline{2-4} &
$\tau \tau$ & $-1.4^{+6.3}_{-5.5}$ & \cite{CMS_tthComb}
\\ \cline{2-4} &
multi-$\ell$ & $3.7^{+1.6}_{-1.4}$ & \cite{CMS_tthML} 
\\ \hline 
inclusive &
$Z \gamma$ & $-0.21 \pm 4.86$ & \cite{CMS_ZGa} 
\\ \cline{2-4} &
$\mu \mu$ & $2.9^{+2.8}_{-2.7}$ & \cite{CMS_mumu} 
\\ \hline 
\end{tabular}

\end{center}
\end{table}

\section{Constraints on Higgs couplings}

We fit the 7 parameters in Eq.~\ref{eq:parameters} to the available Higgs and electroweak precision data assuming the errors are in various channels are Gaussian and uncorrelated, except in the cases where correlations are known (as in the case of 2D likelihood or some electroweak precision data).
We include the uncertainty on the prediction of the SM ggH production cross-section by introducing a nuisance parameter with a Gaussian distribution around the central value. For the LHC at $\sqrt{s}=8$~TeV we take\cite{Heinemeyer:2013tqa} the scale error (+7.2\%, -7.8\%) and the PDF error (+7.5\%, -6.9\%) and add those two linearly.
We obtain the following central values and 68\%~CL intervals for the parameters:
\bea &
c_V = 1.04^{+0.02}_{-0.02},
\quad
c_u = 1.27^{+0.35}_{-0.39}, \quad c_d = 1.08^{+0.17}_{-0.26}, \quad c_l = 1.06^{+0.20}_{-0.20},
& \nn &
 c_{gg} = -0.0012^{+0.0016}_{-0.0015}, \quad
c_{\gamma \gamma} = 0.00065^{+0.00093}_{-0.00066}, \quad c_{Z \gamma} = 0.007^{+0.014}_{-0.034}.
\eea
We find $\chi_{\rm SM}^2 - \chi_{\rm min}^2 = 3.1$ which means that the SM gives a perfect fit to the Higgs and electroweak precision data.
When quoting the confidence regions above we ignored degenerate minima of the likelihood function isolated from the SM point where a large 2-derivative Higgs coupling conspires with the SM loop contributions to produce a small shift of the Higgs observables.
Remarkably, the current data already put meaningful limits on {\em all} 7 parameters.
The strong constraint on $c_V$ is dominated by electroweak precision observables, and can be relaxed in the presence of additional tuned contributions to the S and T parameter that could arise from integrating out heavy new physics states.
Ignoring the electroweak precision data in the fit one obtains the weaker constraint $c_V = 1.03^{+0.08}_{-0.08}$.

\begin{figure}[h!]
    \makebox[\textwidth][c]{
        \subfigure[]{\label{subfig:cu_cgg} \includegraphics[width=0.50\textwidth]{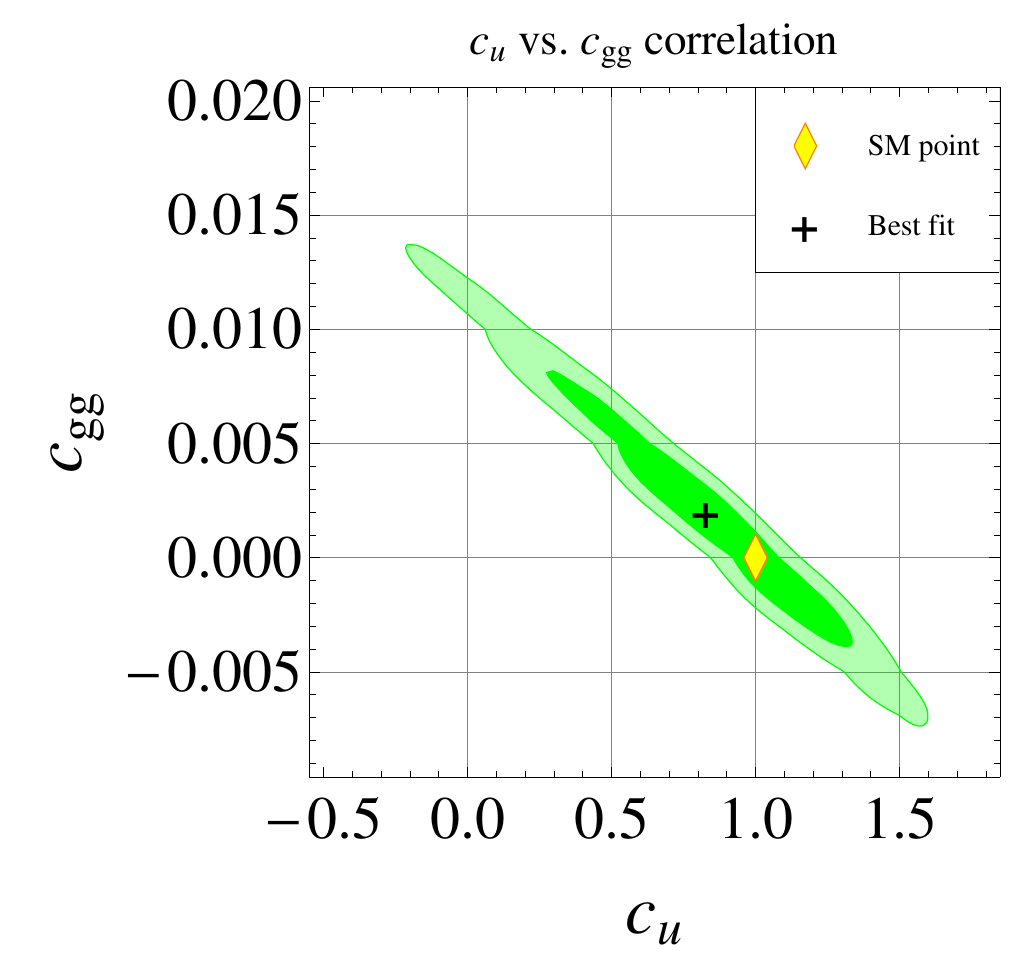}} \hspace{0.1cm}
        \subfigure[]{\label{subfig:BrInv}  \raisebox{0.4cm}{\includegraphics[width=0.60\textwidth]{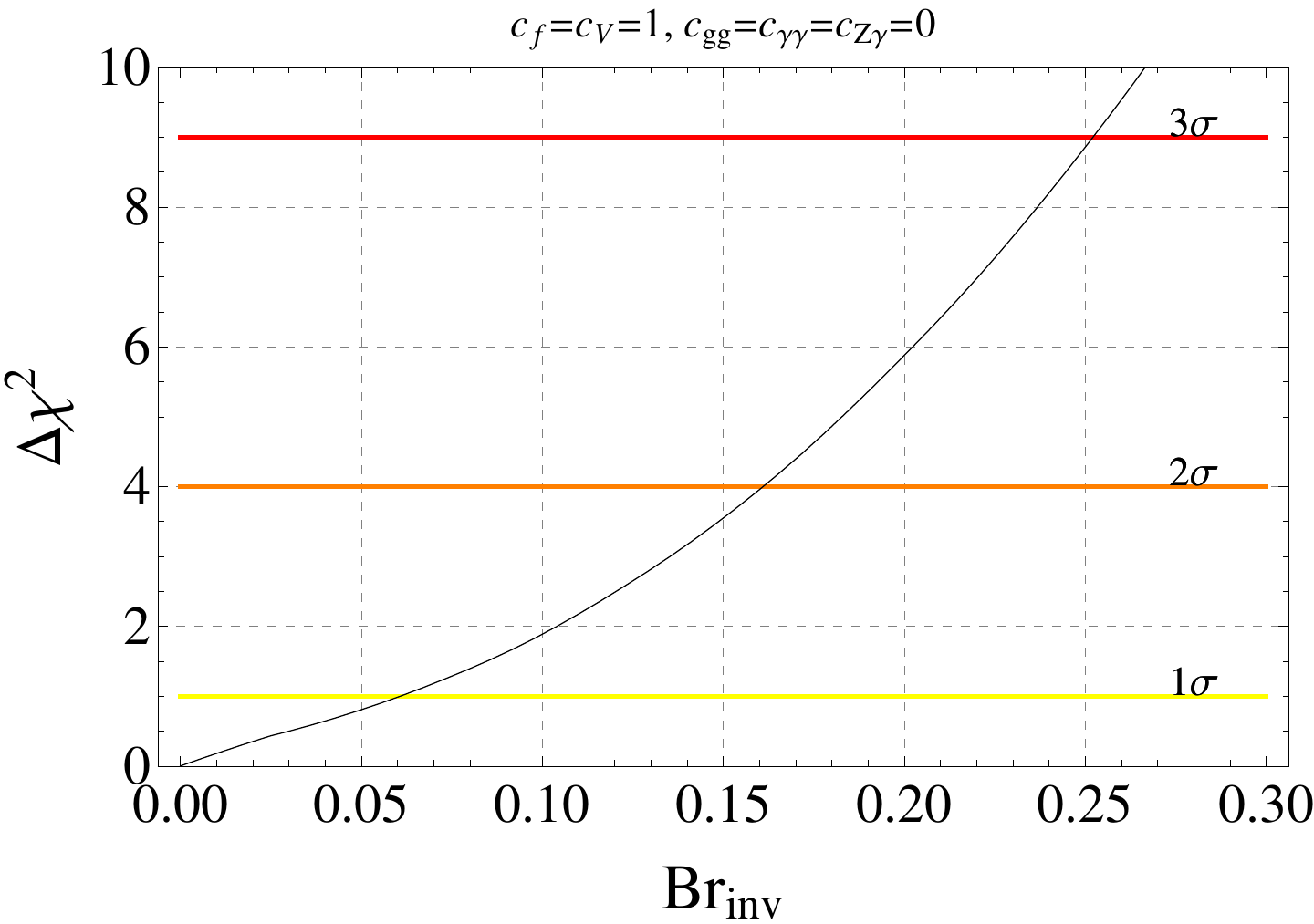}}}
    }
    \caption{Left: A fit in the $c_u$-$c_{gg}$ plane with the other couplings fixed at their Standard Model values. Right: $\chi^2-\chi^2_{min}$ of the fit for the Higgs with SM-size couplings to the SM matter and an invisible branching fraction.}
\end{figure}

The fit displays an approximately flat direction along $c_{gg} + 0.013 c_u$, which is the combination that sets the strength of the gluon fusion production mode. This is clearly visible in Fig.~\ref{subfig:cu_cgg} where a 2D fit in the $c_u$--$c_{gg}$ plane is performed, with the other couplings fixed to their SM values. This flat direction is lifted by the ttH production mode which depends on $c_u$ only.
The recent CMS results in the ttH channel~\cite{CMS_tthComb} put interesting constraints on $c_u$ independently of $c_{gg}$: unlike in the previous fits, $c_u =0$ is now disfavored at the $2\sigma$ level. Furthermore, the fit shows a strong preference for $c_d \neq 0$ even though the $h \to b \bar b$ decay has not been clearly observed. The reason is that $c_d$ determines $\Gamma_{bb}$ which dominates the total Higgs decay width and the latter is indirectly constrained by the Higgs rates measured in other decay channels.
The least stringent constraint is currently that on $c_{Z \gamma}$ which reflects weak experimental limits on the $h \to Z \gamma$ decay rate.
It is interesting to note that there are good prospects~\cite{Chen:2013ejz} of probing $c_{Z \gamma}$ using differential cross section measurements in the $h \to ZZ^* \to 4 \ell$ channel. 

\section{Constraints on invisible width}

Going beyond the effective Higgs Lagrangian in Eq.~\ref{eq:leff}, it is interesting to consider the possibility of an invisible Higgs width. 
This may arise in models with new weakly interacting light degrees of freedom that have a significant couplings to the Higgs boson, for example in Higgs-portal models of dark matter or in supersymmetric models. 
The invisible decays have been directly searched for at the LHC.
The current 95\%~CL limits on the invisible branching fraction are ${\rm Br}_{\rm inv} < 65\%$ in the ZH production mode in ATLAS \cite{ATLAS_invzh}, ${\rm Br}_{\rm inv} < 75\%$ in the ZH production mode in CMS \cite{CMS_invzh}, and ${\rm Br}_{\rm inv} < 69\%$ in the VBF production mode in CMS \cite{CMS_invvbf}.
Stronger limits on the invisible Higgs width can be obtained indirectly from a global fit to the Higgs couplings. 
In the case when the couplings of the Higgs to the SM matter take the SM values the invisible width leads to a universal reduction of the decay rates in all the visible channels. 
This possibility is strongly constrained, given the Higgs is observed in several channels with the rate close to the SM one. From Fig.~\ref{subfig:BrInv} one can read off the limit ${\rm Br}_{\rm inv} < 16\%$ at 95\%~CL.
This bound can be relaxed if one allows new physics to modify the Higgs couplings such that the Higgs production cross-section is enhanced, so as to offset the reduction of the visible rates.
For example, if $c_{gg}$ is allowed to float freely in the fit, the weaker limit ${\rm Br}_{\rm inv} < 40\%$ is obtained. 
Note that these indirect limits apply to any other exotic (but not necessarily invisible) contribution to the Higgs width.

\section*{Acknowledgments}
AF thanks Dean Carmi, Erik Kuflik, Francesco Riva, Alfredo Urbano, Tomer Volansky and Jure Zupan for collaboration on closely related projects. AF also thanks the organizers of the conference {\it Windows on the Universe} for the invitation and support.

\end{document}